\documentclass[intlimits,twoside,a4paper]{article}

\usepackage{amsmath,amssymb}
\usepackage{graphicx}

\usepackage[T2A]{fontenc}
\usepackage[cp1251]{inputenc}
\usepackage[ukrainian,english]{babel}
\usepackage{color}
\def\ds{\displaystyle}

\newcommand\be{\begin{equation}}
\newcommand\ee{\end{equation}}
\newcommand\bee{\begin{eqnarray}}
\newcommand\eee{\end{eqnarray}}

\usepackage[eqsecnum]{cmpj2}

\issue{2017}{20}{1}{13604}
\doinumber{10.5488/CMP.20.13604}
\title[Behavior of the impurity atom in a weakly-interacting Bose gas]%
{Behavior of the impurity atom in a weakly-interacting Bose gas%
}
\author[G.~Panochko, V.~Pastukhov, I.~Vakarchuk]{ G.~Panochko\refaddr{label1},
        V.~Pastukhov\refaddr{label2}, I.~Vakarchuk\refaddr{label2}}
\addresses{
\addr{label1} College of Natural Sciences, Ivan Franko National University of Lviv, 107 Tarnavsky St., 79010 Lviv, Ukraine
\addr{label2} Department for Theoretical Physics, Ivan Franko National University of Lviv,\\ 12 Dragomanov St., 79005 Lviv, Ukraine
}

\authorcopyright{G.~Panochko, V.~Pastukhov, I.~Vakarchuk, 2017}
\date{Received January 16, 2017, in final form February 27, 2017}
\begin{document}

\maketitle

\begin{abstract}
We studied the properties of a single impurity atom immersed in a dilute Bose condensate at low temperatures. In particular, we  perturbatively obtained the momentum dependence of the impurity spectrum and damping. By means of the Brillouin-Wigner perturbation theory we also calculated the self-energy both for attractive and repulsive polaron in the long-wavelength limit. The stability problem of the impurity atom in a weakly-interacting Bose gas is also examined.
\pacs 67.60.Bc, 03.75.Hh

\end{abstract}

\section{Introduction}

When an impurity atom is immersed in a dilute Bose gas, impurity-boson interactions are expected to dress the impurity into a quasiparticle, namely a Bose polaron. Methods for obtaining the dynamical properties of such a quasiparticle are the subject of research in many papers.

Investigation of moving impurities in Bose systems was motivated by the progress in the experimental research, where charged or localized impurities in a Bose-Einstein condensate \cite{Schmid} and impurities interacting with uncondensed bosonic medium \cite{Spethmann} were obtained.
In recent work \cite{Hu}, the lifetime of a Bose polaron in the ultracold $^{87}$Rb Bose gas in a strongly interacting regime was experimentally estimated. In \cite{Jorgensen}, the dependence of the  energy of impurity atom which is moving in a Bose-Einstein condensate on the variation of the interaction between the impurity atom and the bosonic system is measured.

A lot of theoretical works, where the properties of a Bose polaron are investigated by various methods, have been published in recent years. Namely, attractive and repulsive interaction between the impurity atom and the environment is considered. In particular, the self-localised impurity states are studied in a one-dimensional \cite{Sacha} and in a dilute three-dimensional Bose-Einstein condensate \cite{Cucchietti} in a strong coupling regime. The strongly interacting Bose polaron is also investigated in \cite{Rath}. By means of
the self-consistent T-matrix approach, the authors calculated the impurity spectral function, the effective mass, the quasiparticle energy of the attractive and repulsive Bose polarons. Qualitatively, the same results were obtained in \cite{Li} with the help of variational methods. Particularly, the parameters of the spectrum of the impurity atom immersed in the non-interacting BEC and in a Bose gas with weak repulsive interaction between particles were found. The systematic diagrammatic techniques applied in \cite{Christensen} to the polaron problem revealed a similarity between the structure of polaron energy and the energy of weakly-interacting Bose gas.  In the paper \cite{Shchadinova},
the dynamics of ultracold Bose polaron was analyzed and smooth transition from the attractive polaron to a molecular state near the Feshbach resonance was explained. The quantum Monte Carlo method was used in  \cite{Pena1,Pena2} to study the dependence of the Bose polaron energy on the gas parameter at different mass ratios of the impurity atom and Bose particle.
In the work \cite{Sun}, the impurity self-energy and the spectral function for $^{6}$Li--$^{133}$Cs system was determined by the diagrammatic approach considering that the impurity can form a sequence of the Efimov bound states with two bosons. In \cite{Shashi}, using the Lee-Low-Pines transformation, the absorption spectra of impurity with two or more internal hyperfine states is calculated.

In this paper, we consider the behavior of the impurity atom in a weakly-interacting Bose gas. We focus our attention on finding the impurity spectrum in the Bogoliubov approach. We have found the correction to the impurity spectrum within the standard perturbation theory, assuming that the Bose particles are hard spheres (see section~\ref{sec:Sp}). The most interesting results for
the self-energy and the effective mass of the attractive and repulsive polarons were obtained within the Brillouin-Wigner perturbation theory.
We have also showed that the spectrum of an impurity is damped in the case of a positive value of the Bose polaron energy and calculated the dependence of the damping in the limit of the Rayleigh-Schr\"odinger perturbation theory on the velocity of the impurity atom (see section~\ref{sec:Sps1}). Considering various values for the interaction between bosons and the impurity, as well as the gas parameter values, we were able to find the region of applicability for our approach.

\section{Problem statement}\label{sec:Problem}
The system $N$ of interacting Bose particles which also interacts with the impurity atom  can be described by the Hamiltonian:
\be\label{21}
\widehat{H} = \widehat{H}_{\text I}+\widehat{H}_{\text L}+\widehat{H}_{\text{int}}.
\ee
Here, $\widehat{H}_{\text I}$ is the kinetic energy of the impurity:
\be\label{22}
\widehat{H}_{\text I}=\frac{\widehat{P}^2}{2M}.
\ee
The operator $\widehat{H}_{\text L}$ is the Hamiltonian of the interacting Bose particles. In Bogoliubov approximation, this Hamiltonian  has a diagonal form in the second quantization representation $\widehat{b}_{\mathbf{k}}^{+}$ $\widehat{b}_{\mathbf{k}}$:
\be\label{25}
\widehat{H}_{\text L}=E_{\text B}+\sum_{\mathbf{k}\neq 0}\hbar\omega_k\widehat{b}_{\mathbf{k}}^{+}\widehat{b}_{\mathbf{k}}\,,
\ee
where $E_{\text B}$ is the ground-state energy of the Bose particles in the Bogoliubov's approximation, $\hbar\omega_k$ is the Bogoliubov's spectrum.
The operator $\widehat{H}_{\text{int}}$ describes the interaction between bosons and impurity with two-body scattering processes of the impurity atom on the Bose particles taken into account:
\be\label{26}
\widehat{H}_{\text{int}}=\rho\bar{\nu}_0+\ds\frac{1}{\sqrt{V}}\sum_{\mathbf{k}\neq 0}\sqrt{\rho}\bar{\nu}_k\frac{\hbar k^2}{2m \omega_k}\left(
\widehat{b}_{\mathbf{-k}}^{+}\re^{\ri\mathbf{kr}}+\widehat{b}_{\mathbf{k}}\re^{-\ri\mathbf{kr}}\right),
\ee
where $\rho=N/V$ and $m$ is the density and mass of Bose particles, respectively. The coordinate of the impurity atom $\mathbf{r}$ is introduced. The coefficients $\bar{\nu}_0$ and $\bar{\nu}_k$  are the Fourier transforms of the interaction potential between the impurity and Bose particles when $k=0$ and $k\neq0$, respectively.

To describe the interaction between the bosons and the impurity atom we can use the model potential of hard  spheres instead of the real potential. This potential gives an adequate description of the contact repulsion between particles. The Fourier image
of this pseudopotential is constant for all the values of the wave vectors:
\be\label{27}
\bar{\nu}_k=2\pi\hbar^2\bar{a}\frac{M+m}{mM}\,,
\ee
and is determined by the $s$-scattering length $\bar{a}$ of the impurity atom on a Bose particle. The moving impurity in the environment of weakly interacting Bose particles is usually called the Bose polaron. Moreover, the sign of the coupling constant $\bar{\nu}_k$ in the relation  (\ref{27})
indicates whether the polaron is attractive ($\bar{\nu}_k<0$) or repulsive ($\bar{\nu}_k>0$).

Our task is to find the full energy of the system ``impurity atom plus Bose particles'' by using the Brillouin-Wigner perturbation theory. We will also consider the limit of the Rayleigh-Schr\"odinger perturbation theory.

Let the impurity atom with momentum $\hbar \mathbf{q}$ move in the environment of the bosons, which are in the ground state
$|0\rangle$ with the energy $E_{\text B}$. The full energy of that system can be written in the form of  transcendental equation:
\be\label{28}
E_q=E_q^{(0)}+\rho\bar{\nu}_0-I(E_q),
\ee
where $E_q^{(0)}$ is  the zeroth approximation of the ground state energy of the system ``the impurity atom plus interacting bosons'':
\be\label{29}
E_q^{(0)}=\frac{\hbar^2q^2}{2M}+E_{\text B}\,,
\ee
the second term in (\ref{28}) is the diagonal matrix element of the perturbation operator (\ref{26}). It is calculated on the ground state wave functions of the system ``the impurity atom plus Bose particles''.  The next term in (\ref{28}) is second order correction to the full energy:
\be\label{210}
I(E_q)=\mathop{\sum_{\mathbf{q'}\neq0,\,\,\mathbf{k'}\neq0,}}\limits_{\mathbf{q'}+\mathbf{k'}\neq 0}
\frac{|\langle\mathbf{k'},\mathbf{q'}|\widehat H_{\text{int}}|0,\mathbf{q}\rangle|^2}{E^{(0)}_{q',k'}-E_q}\,.
\ee
Since we treat the Bose gas as a weakly-interacting gas, the most probable transitions in this system are those which are accompanied by the appearance of a single phonon.
An intermediate state $|\mathbf{k'},\mathbf{q'}\rangle$ can be determined by the momentum of quasiparticles $\hbar \mathbf{q'}$ and the energy of the first excited state of the environment:
\be\label{211}
E^{(0)}_{q',k'}=\frac{\hbar^2q'^2}{2M}+\hbar\omega_{k'}+E_{\text B}\,.
\ee
In order to find the energy $E_q$, it is convenient to perform an analytic continuation in the complex energy plane $E_q\to E_q+\ri\eta$ (here $\eta \rightarrow +0$). In what follows we will continue to work with its real and imaginary parts:
\be\label{212}
E_q-E_{\text B}\equiv \Delta \varepsilon_q-\ri\Gamma_q\,,
\ee
where $\Delta \varepsilon_q$, $\Gamma_q$ are determined by the second-order correction (\ref{210}) of the perturbation theory. The real and imaginary parts of this correction can be found throughout the Sokhotski formula.
So,
\be\label{213}
\Delta \varepsilon_q=\frac{\hbar^2q^2}{2M}+\rho\bar\nu_0-\Re I(\Delta\varepsilon_q),
\ee
\be\label{214}
\Gamma_q=\Im I(\Delta\varepsilon_q).
\ee
The impurity atom with momentum $\hbar \mathbf{q}$ always loses its energy $\Delta
\varepsilon_q$ when it moves in the environment formed by Bose particles. Even when the impurity atom is decelerated by weakly-interacting bosons,  its energy is not equal to zero.  This immersion energy which appears due to the interaction with Bose particles is usually called the self-energy of the impurity atom. If the sign of this interaction is negative, then the energy $\Delta\varepsilon_q$ can attain different values. If the impurity atom is repulsed by the environment of the Bose particles, then $\Delta\varepsilon_q>0$ and this energy is converted into the emission of phonons.  Moreover, in the case of a strong attractive impurity-boson interaction, various phenomena can occur, namely,  self-localization of the impurity or even the formation of the polaron-boson bound states ``impurity plus Bose particle''. The imaginary part of the spectrum $\Gamma_q$ determines the rate of the energy loss of the impurity and indicates the lifetime of the polaron quasiparticle. Therefore, the immersion energy of the impurity atom should be much greater than the magnitude of the spectrum attenuation $\Gamma_q$. The impurity atom almost immediately loses its energy and slows down when $\Delta \varepsilon_q<\Gamma_q$. It should be also noted that for the case $\Delta\varepsilon_q\backsimeq\Gamma_q$, equations~(\ref{213}) and (\ref{214}) are a system of transcendental equations in which $I(\Delta\varepsilon_q \to \Delta\varepsilon_q-\ri\Gamma_q )$. The solution of this problem is quite cumbersome. Therefore, we will consider the case $\Delta\varepsilon_q\gg\Gamma_q$ and then discuss the applicability of our approach.

\section{The impurity spectrum}\label{sec:Sp}
Let us write the impurity spectrum (\ref{213}) as follows:
\be\label{41}
\Delta\varepsilon_q=\frac{\hbar^2q^2}{2M}+\rho\bar{\nu}_0
-\frac{\rho\bar{\nu}_{0}^2}{V}\sum_{\mathbf{k'}\neq 0}
\ds\ds\frac{\hbar^2k{'}^2}{2m}\frac{1/\hbar\omega_{k'}}{\frac{\hbar^2 k'^2}{2M}+\frac{\hbar^2
		\mathbf{k'q}}{M}+\hbar\omega_{k'}-\Delta\varepsilon_q}\,,
\ee
here, Bogoliubov's spectrum can be expressed through the Fourier transform of the interaction potential between Bose particles $\nu_0$:
\be\label{411}
\hbar\omega_{k'}=\ds\frac{\hbar^2k'^2}{2m}\sqrt{1+2\rho\nu_k\left(\frac{\hbar^2k'^2}{2m}\right)^{-1}}\,.
\ee
The last term in  (\ref{41}) is divergent because we describe the interaction between the impurity and bosons by the model potential.
The convergence of this contribution is provided by renormalization of the coupling constant $\bar{\nu}_0$ \cite{Vakarchuk,Fetter}:
\be\label{42}
\bar{\nu}_{0} \to \bar{\nu}_{0}+\ds\frac{\bar{\nu}_{0}^2}{V}\sum_{\mathbf{k'}\neq 0}\ds\frac{2Mm}
{(m+M)\hbar^2k^2}\,.
\ee
In the case of more realistic interaction potential between the impurity atom and the Bose system, this convergence is provided by the Fourier transform $\bar\nu_{k}\neq \text{const}$ which tends to zero for large values of $\mathbf{k'}$ while it is constant in the long-wavelength limit.

We write down the damping of spectrum of the impurity atom (\ref{214}) in an explicit form:
\be\label{31}
\Gamma_q=\pi\ds\frac{\rho\bar{\nu}_{0}^2}{V}\sum_{\mathbf{k'}\neq 0}
\frac{\hbar k'^2}{2m\omega_{k'}}
\ds\delta\left(\frac{\hbar^2 k'^2}{2M}+\hbar\omega_{k'}+\frac{\hbar^2 \mathbf{k'q}}
{M}-\Delta\varepsilon_q\right).
\ee
The integral in expression (\ref{31}) for the damping in the case of arbitrary interaction (when the Fourier transform of the interaction potential is constant and when $\bar{\nu}_k\neq \text{const}$) is always convergent. Moreover, due to the presence of $\delta$-function, it is also clear that $\Gamma_q$ is not equal to zero only when the self-energy of the impurity is positive, i.e., $\Delta\varepsilon_q>0$.

In next sections we will investigate the full momentum dependence of the impurity spectrum and damping within Rayleigh-Schr\"odinger second-order perturbation theory and will study only the long-wavelength behavior of these parameters by means of Brillouin-Wigner perturbative approach.

\subsection{The case of the Rayleigh-Schr\"odinger perturbation theory}\label{sec:Sps1}

\begin{figure}[!t]
\vspace{-3mm}
\centering
\includegraphics[scale=0.28]{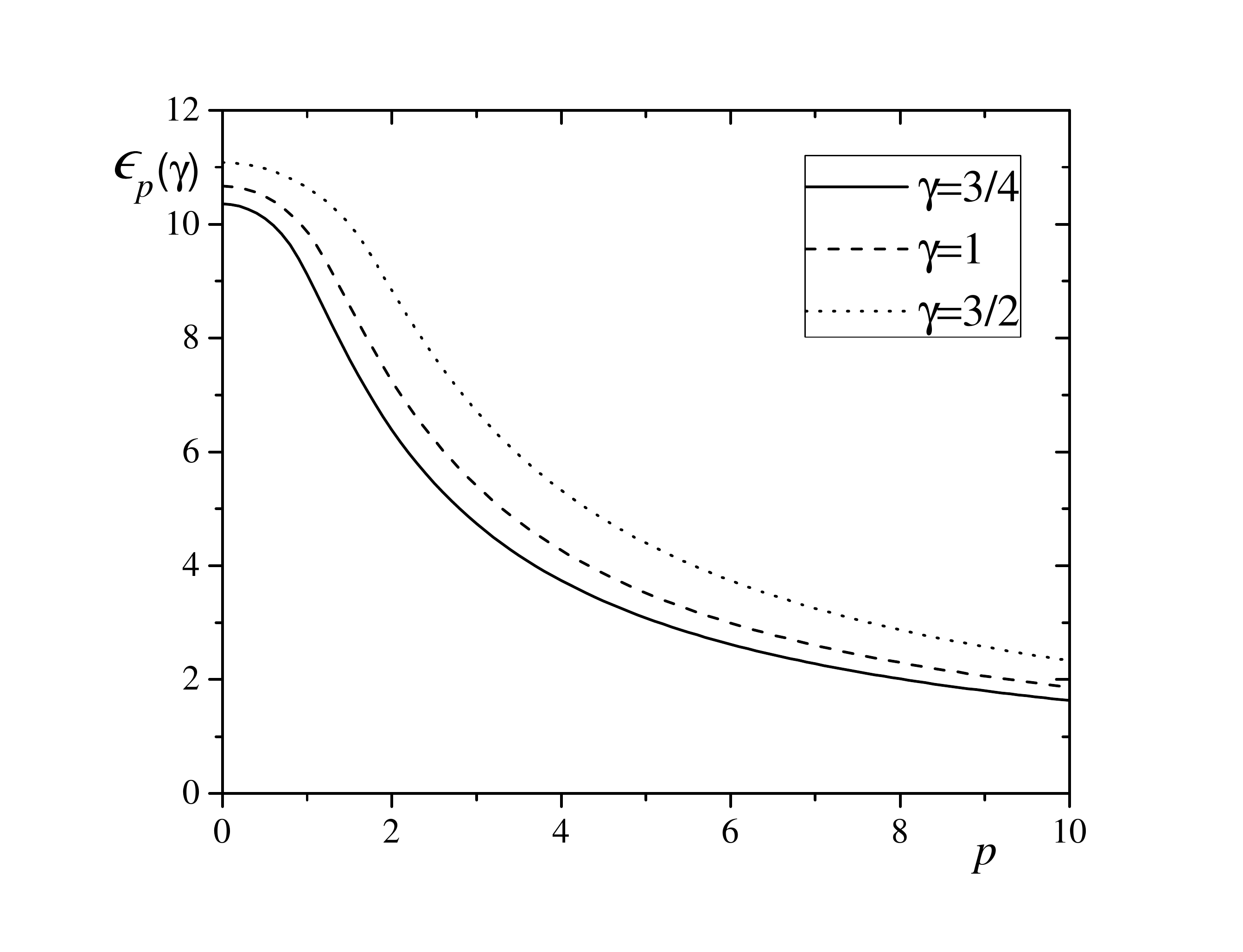}\includegraphics[scale=0.28]{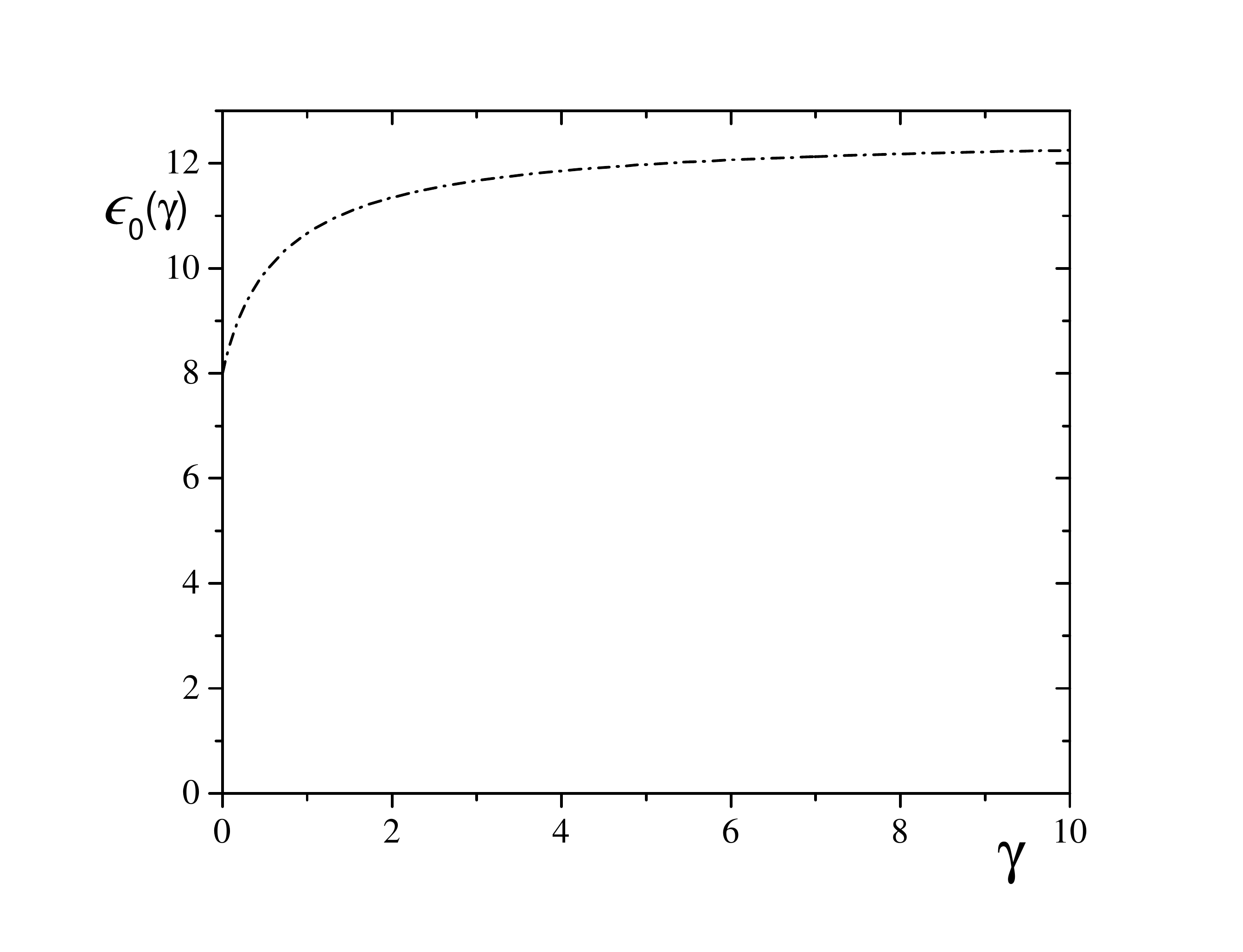}
\vspace{-5mm}
		\caption{The correction of the impurity spectrum in the limit of the
		Rayleigh-Schr\"odinger perturbation theory.
		Left-hand: the dependence of the corrections on momentum of the impurity atom. Right-hand: the dependence of the corrections on mass of the impurity in the region of small momenta. \label{fig00}}
\end{figure}

The calculation of the spectrum of the impurity atom in the limit of the standard perturbation theory, when $\Delta\varepsilon_q=0$ in the last term equation~(\ref{41}), is rather simple. In the thermodynamic limit ($V\to \infty$, $N\to\infty$, $\rho =\text{const} $), where
$\sum_{\mathbf{k'}}\to\frac{V}{(2\pi)^3}\int \rd\mathbf{k'}$,
introducing a dimensionless momentum of the impurity $p=\frac{\hbar q}{mc}$
(here $c=\sqrt{\rho{\nu}_0/m}$ is the velocity of sound in the environment of the Bose particles) and calculating the integral over the variable ${\mathbf k}'$ in the sense of principal value, we obtain the final expression for the spectrum of a Bose polaron:
\be\label{43}
\Delta\tilde{\varepsilon}_p=\ds\frac{p^2}{2\gamma}+
\frac{\bar{a}}{a}\frac{1+\gamma}{2\gamma}\left[1+\frac{\bar{a}}{a}\sqrt{\frac{\rho a^3}{\pi}}
\epsilon_p(\gamma)\right],
\ee
here, $\gamma=M/m$, $\Delta\tilde{\varepsilon}_p=\Delta\varepsilon_p/mc^2$ are the impurity dimensionless mass and energy, respectively.  For the explicit form of the function $\epsilon_p(\gamma)$ see appendix~\ref{sec:Apa}.
The parameter $\rho a^3$ is the gas parameter
which determines the strength of the short-range repulsion between Bose particles. In figure~\ref{fig00}, we present the results of numerical calculations of the function $\epsilon_p(\gamma)$. The correction to the spectrum $\epsilon_p(\gamma)$ essentially depends on the mass ratio $\gamma$.

In the limit of $p\to \infty$, the leading term in (\ref{43}) is quadratic in momentum. The correction to the spectrum tends to a constant when $p\to 0$ (see appendix~\ref{sec:Apa}). So, we conclude that even the immobile impurity due to interaction with Bose system will have a nonzero energy.

The sum over ${\mathbf k}'$ in equation~(\ref{31}) was calculated in the thermodynamic limit when $\Delta\varepsilon_q=0$. Taking into account the argument of the $\delta$-function which is nothing but the energy conservation in the single phonon creation process, it is easily seen
that the impurity atom with momentum $\hbar\mathbf{q}$ will transfer its energy to the bosons when the velocity of the impurity atom
is greater than the velocity sound of the environment. This fact should be taken into account when calculating the integrals in equation~(\ref{31}). Omitting the details of computations, we write down the final expression for the damping of spectrum in dimensionless variables:
\be\label{34}
\tilde{\Gamma}_p=\sqrt{\pi\rho a^3}\ds\left(\frac{\bar{a}}{a}\right)^2\ds\frac{(1+\gamma)^2}{\gamma}
\frac{1}{p}\left(k_p\sqrt{1+k_{p}^2}-\ln\left|k_p+\sqrt{1+k_p^2}\right|\right),
\ee
where notations
\be\label{35}
k_p=\ds\frac{1}{1-\gamma^2}\left(p-\gamma\sqrt{p^2+1-\gamma^2}\right),\qquad {\Gamma}_p=\Gamma_p/mc^2
\ee
are used. When $\gamma=1$, the damping can be written in a simple analytical form:
\be\label{36}
\tilde{\Gamma}_p=\sqrt{\pi\rho a^3}\ds\left(\frac{\bar{a}}{a}\right)^2
\left(\frac{p^4-1}{p^3}-4\ds\frac{\ln p}{p}\right),
\ee
and here, $p>1$.

Thus, in the limit $\Delta\tilde{\varepsilon}_q=0$, the spectrum damping $\tilde\Gamma_p$ essentially depends on dimensionless momentum of the impurity $p$ when $p>\gamma$. In the case of immobile impurity, the damping is equal to zero. This can be easily seen calculating the limit $\tilde{\Gamma}_{p\to 0}$ in the relation (\ref{34}). In the case $\tilde{\Gamma}_{p\to \infty}\sim p$, the damping increases linearly with increasing the velocity of the impurity, but the condition $\Gamma_p/\Delta\varepsilon_p\ll1$ is satisfied. Finally, taking into account the above calculations, we conclude that the impurity states in a dilute Bose condensate are always well-defined (i.e., $\Gamma_p/\Delta\varepsilon_p\ll1$) in the weak-coupling limit $\sqrt{\rho a^3}(\bar{a}/{a})^2\ll 1$.

\subsection{The case of the Brillouin-Wigner perturbation theory}
Let us find the impurity spectrum when $\Delta\varepsilon_q\neq 0$. To do this, we will expand expression (\ref{41}) into a series in powers of $q$:
\be\label{44}
\Delta\varepsilon_q=\Delta\varepsilon_0+\frac{\hbar^2q^2}{2M^{*}}+O(q^4),
\ee
where
\be\label{45}
\Delta\varepsilon_0=\rho\bar{\nu}_0-\frac{\rho\bar{\nu}_{0}^2}{V}\sum_{\mathbf{k'}\neq 0}\left(
\ds\frac{\hbar^2 k'^2}{2m}\frac{1/\hbar\omega_{k'}}{\frac{\hbar^2 k'^2}{2M}+\hbar\omega_{k'}-\Delta\varepsilon_0}-\frac{1}{\frac{\hbar^2 k'^2}{2M}+\frac{\hbar^2 k'^2}{2m}}\right)
\ee
is the self-energy of the impurity. In particular, the energy $\Delta\varepsilon_0$ can be both positive and negative. Its sign will be determined by the type of interaction (attractive or repulsive) between the impurity atom and the Bose system.
The second term in (\ref{44}) is the kinetic energy of the impurity atom with the effective mass~$M^{*}$:
\be\label{46}
\frac{M}{M^{*}}=1-\frac{1}{3}\frac{\rho\bar{\nu}_{0}^2}{V}\frac{1}{f(\Delta\varepsilon_0)}\sum_{\mathbf{k'}\neq 0}
\ds\frac{\hbar^4 k'^4}{mM}\ds\frac{1/\hbar \omega_{k'}}{\left(\frac{\hbar^2 k'^2}{2M}+\hbar\omega_{k'}-\Delta\varepsilon_0\right)^3};
\ee
here, the function
\be\label{47}
f(\Delta\varepsilon_0)=1+\frac{\rho\bar{\nu}_{0}^2}{V}\sum_{\mathbf{k'}\neq 0}
\ds\frac{\hbar^2 k'^2}{2m}\frac{1/\hbar\omega_{k'}}{\left(\frac{\hbar^2 k'^2}{2M}+\hbar\omega_{k'}-\Delta\varepsilon_0\right)^2}\,.
\ee
It is seen that the effective mass is determined by the self-energy of the impurity. Therefore, first we have to calculate the impurity energy $\Delta\varepsilon_0$ and then compute $M^{*}$.

The integrals in equations~(\ref{45}), (\ref{46}) and (\ref{47}) can be calculated analytically. Performing these simple but nevertheless cumbersome integrations for the dimensionless energy of impurity, we obtain the following equation:
\be\label{481}
\Delta\tilde{\varepsilon}_0=
\frac{\bar{a}}{a}\frac{1+\gamma}{2\gamma}\left[1+\frac{\bar{a}}{a}\sqrt{\frac{\rho a^3}{\pi}}
\epsilon_0(\gamma,\Delta\tilde{\varepsilon}_0)\right],
\ee
where the explicit form of the contributions $\epsilon_0(\gamma,\Delta\tilde{\varepsilon}_0)$ is shown in appendix \ref{sec:Apa}.

\begin{figure}[!b]
\vspace{-4mm}
\centering
	\includegraphics[scale=0.27]{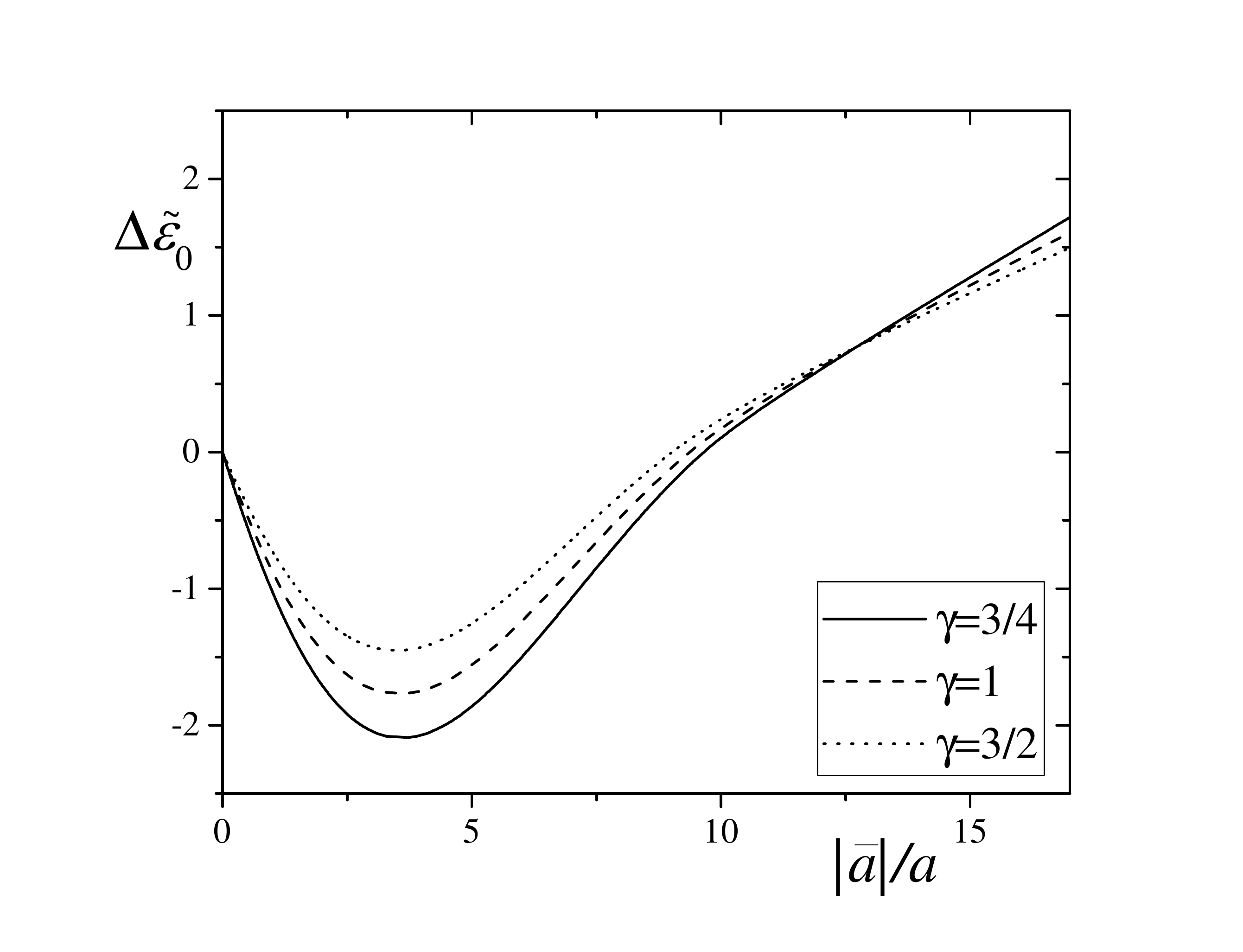}
	\includegraphics[scale=0.27]{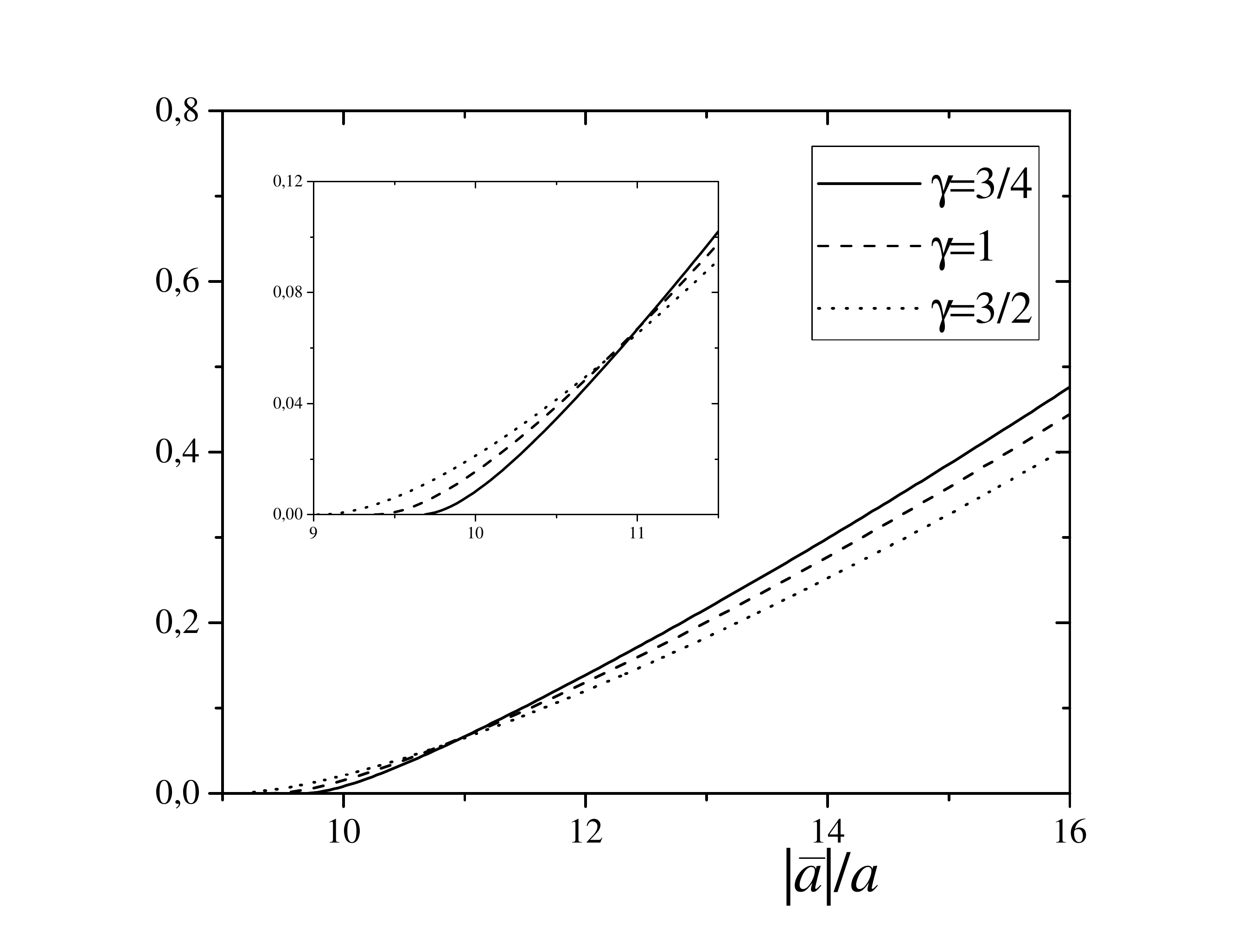}
\vspace{-5mm}\\
	\includegraphics[scale=0.27]{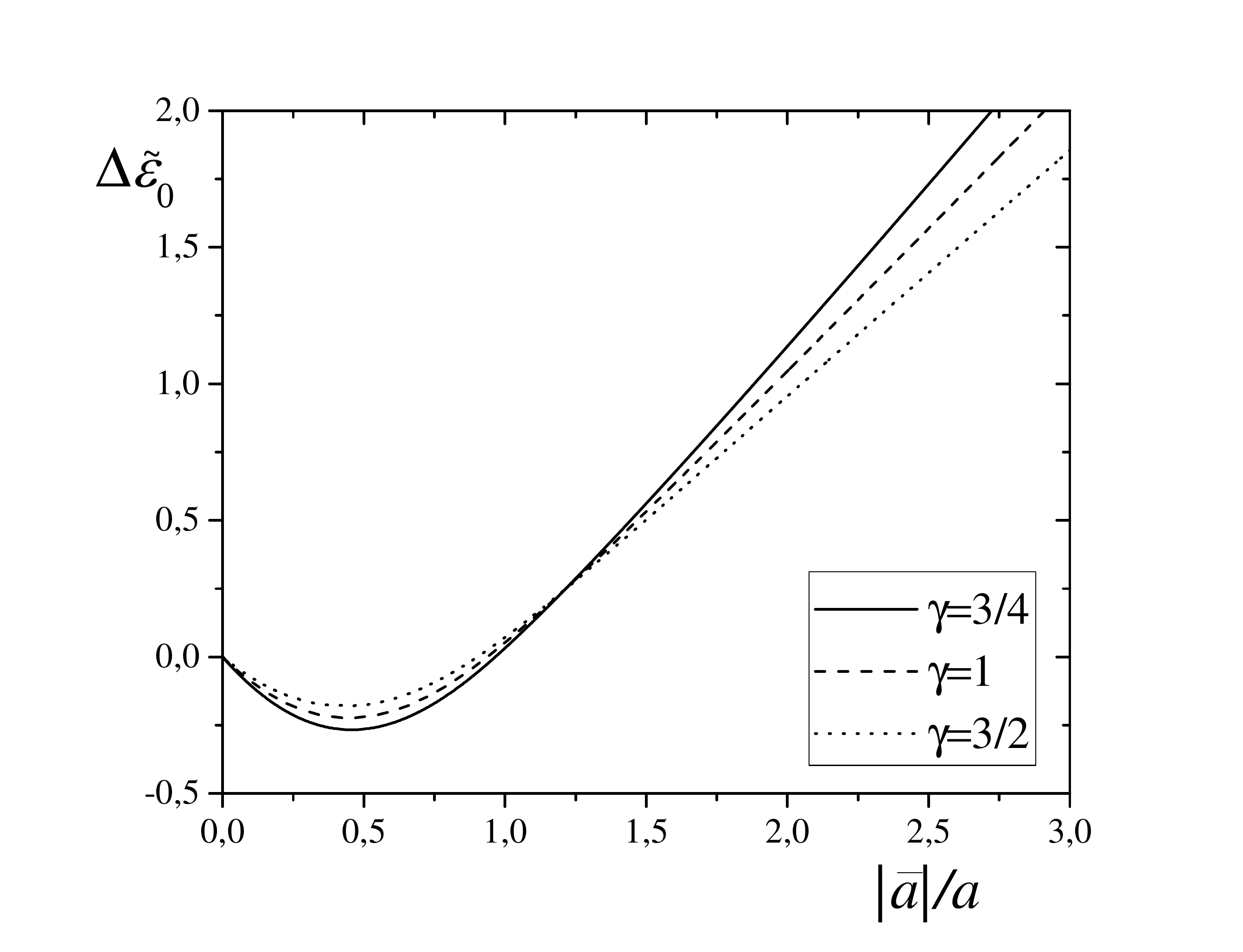}
	\includegraphics[scale=0.27]{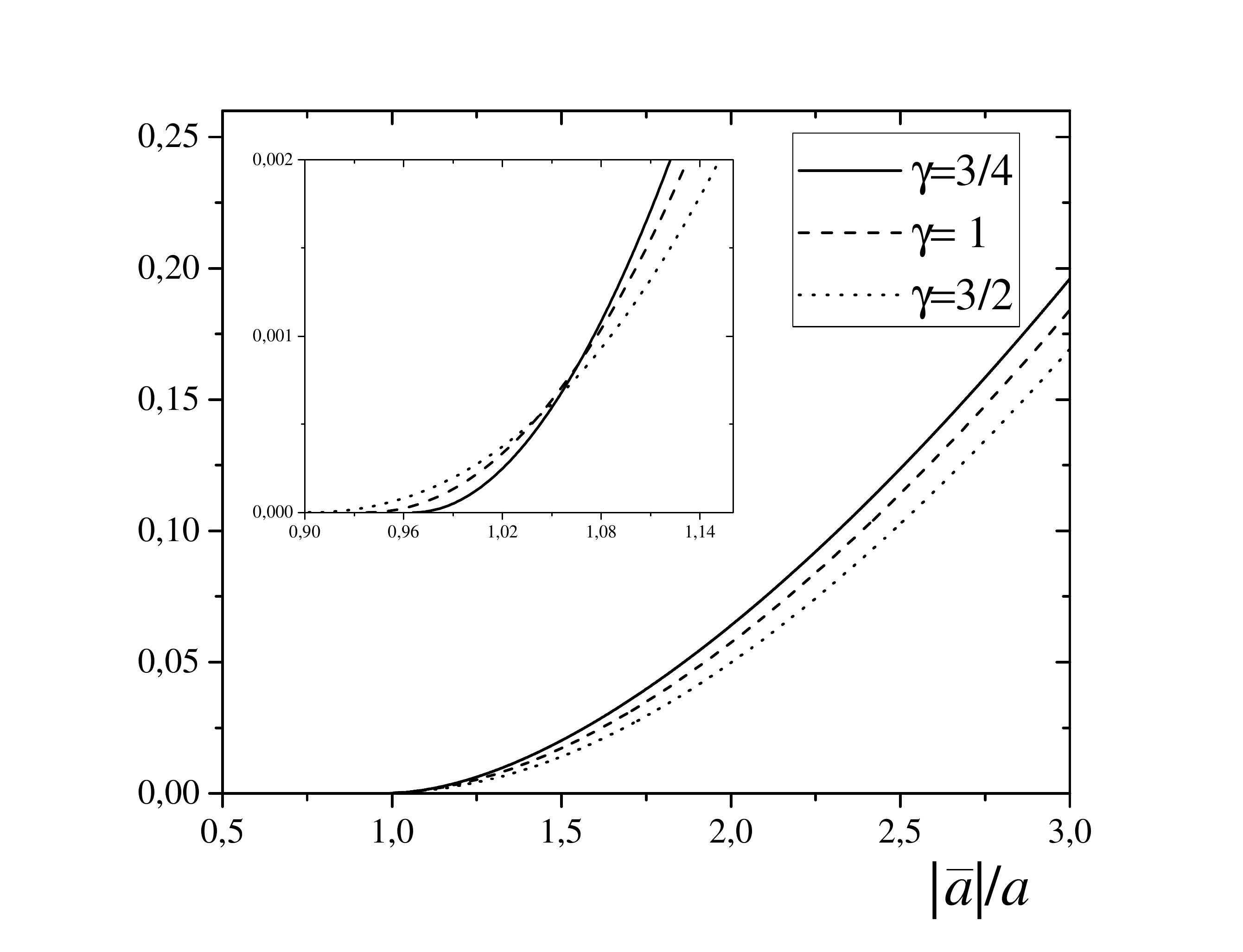}
	\caption{The attractive polaron: left-hand panel is the self-energy of the impurity atom, right-hand panel is the ratio of the spectrum damping to the self-energy impurity atom. (Top row $\rho a^3=5\times10^{-4}$, bottom row $\rho a^3=3\times10^{-2}$). \label{fig0}}
\end{figure}

Despite the standard perturbation theory result where the spectrum damping is always zero in the limit $p\to 0$, the solution of equation~(\ref{214}) could exist even for $\Gamma_0$ in the Brillouin-Wigner prescription. This defect of the approximate consideration of the polaron problem when $\Gamma_0\simeq \Delta\varepsilon_0$, is usually treated as a polaron instability \cite{Rath,Li}. Of course, such a treatment is physically wrong and the magnitude of $\Gamma_0/\Delta\varepsilon_0$ only defines the limits of applicability of the arbitrary approximation scheme. Let us estimate the value of the spectrum damping (\ref{41}) in the limit of the immobile impurity atom with the self-energy $\Delta\varepsilon_0$. After simple calculations, in the dimensionless variables for any values of $\gamma$ we obtain:
\be\label{37}
\tilde{\Gamma}_0=\frac{\pi}{2}\sqrt{\frac{\rho a^3}{\pi}}\left(\frac{\bar{a}}{a}\right)^2\frac{(1+\gamma)^2}{\gamma}\frac{k_0^{3/2}}{\gamma(1+2k_0)+2\sqrt{k_0(1+k_0)}}\,,\qquad \tilde{\Gamma}_0={\Gamma}_0/mc^2,
\ee
where
\bee\nonumber
k_0=\frac{\gamma}{2(1-\gamma^2)}\left[\gamma+\Delta\tilde{\varepsilon}_0-
\sqrt{(\gamma+\Delta\tilde{\varepsilon}_0)^2+\Delta\tilde{\varepsilon}_0^2(\gamma^2-1)}\right].
\eee

\begin{figure}[!b]
\vspace{-4mm}
\centering
	\includegraphics[scale=0.27]{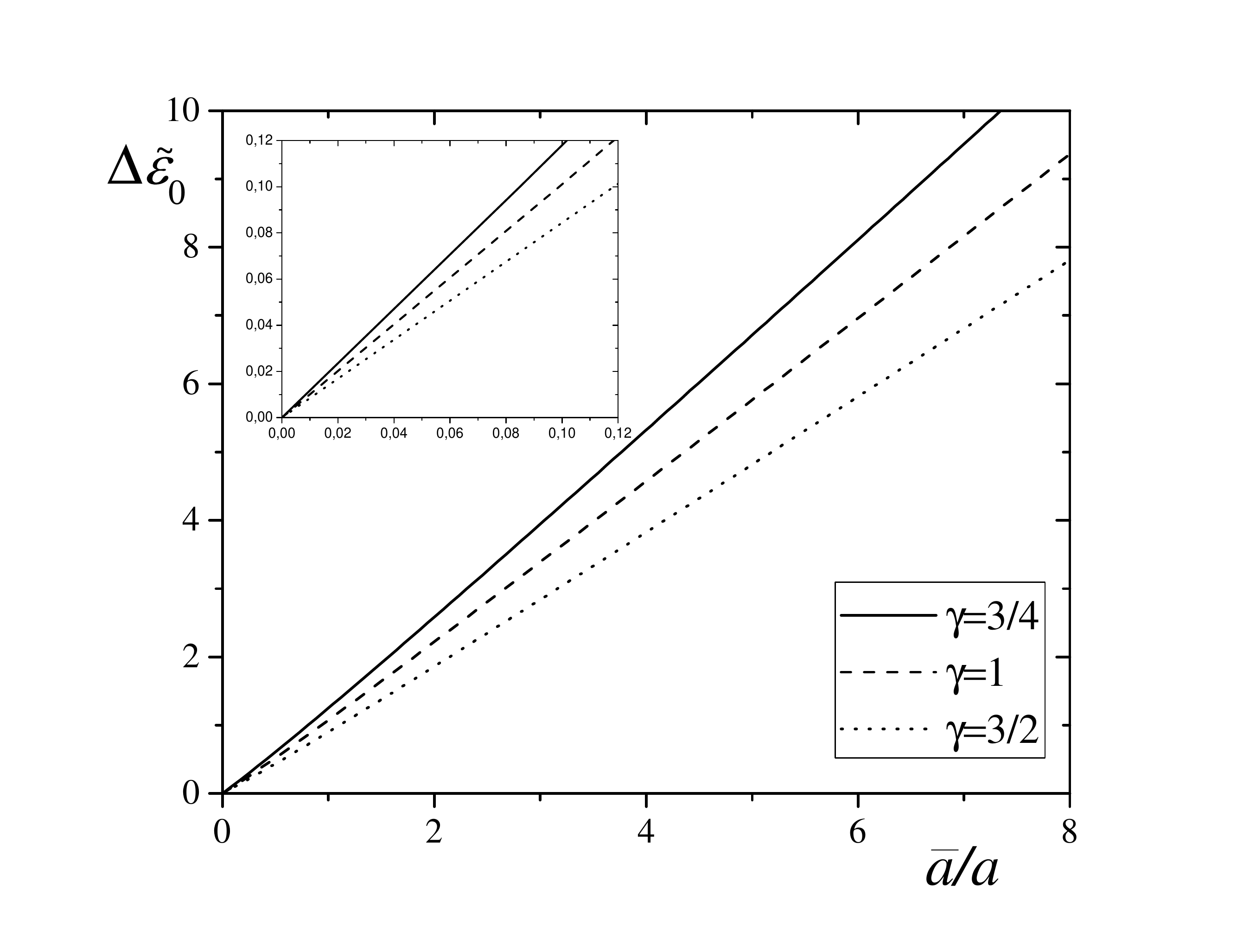}
	\includegraphics[scale=0.27]{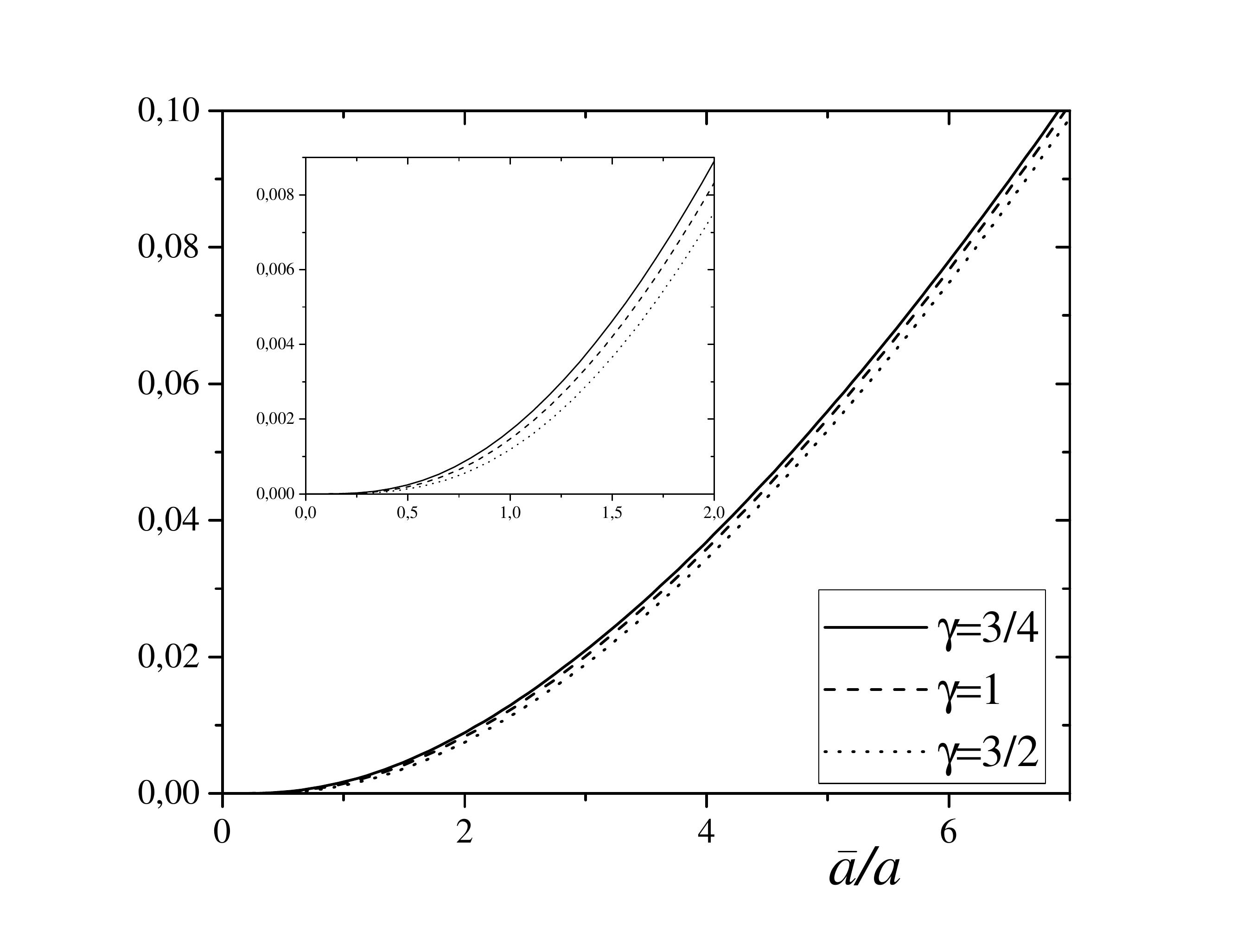}
\vspace{-5mm}\\
	\includegraphics[scale=0.27]{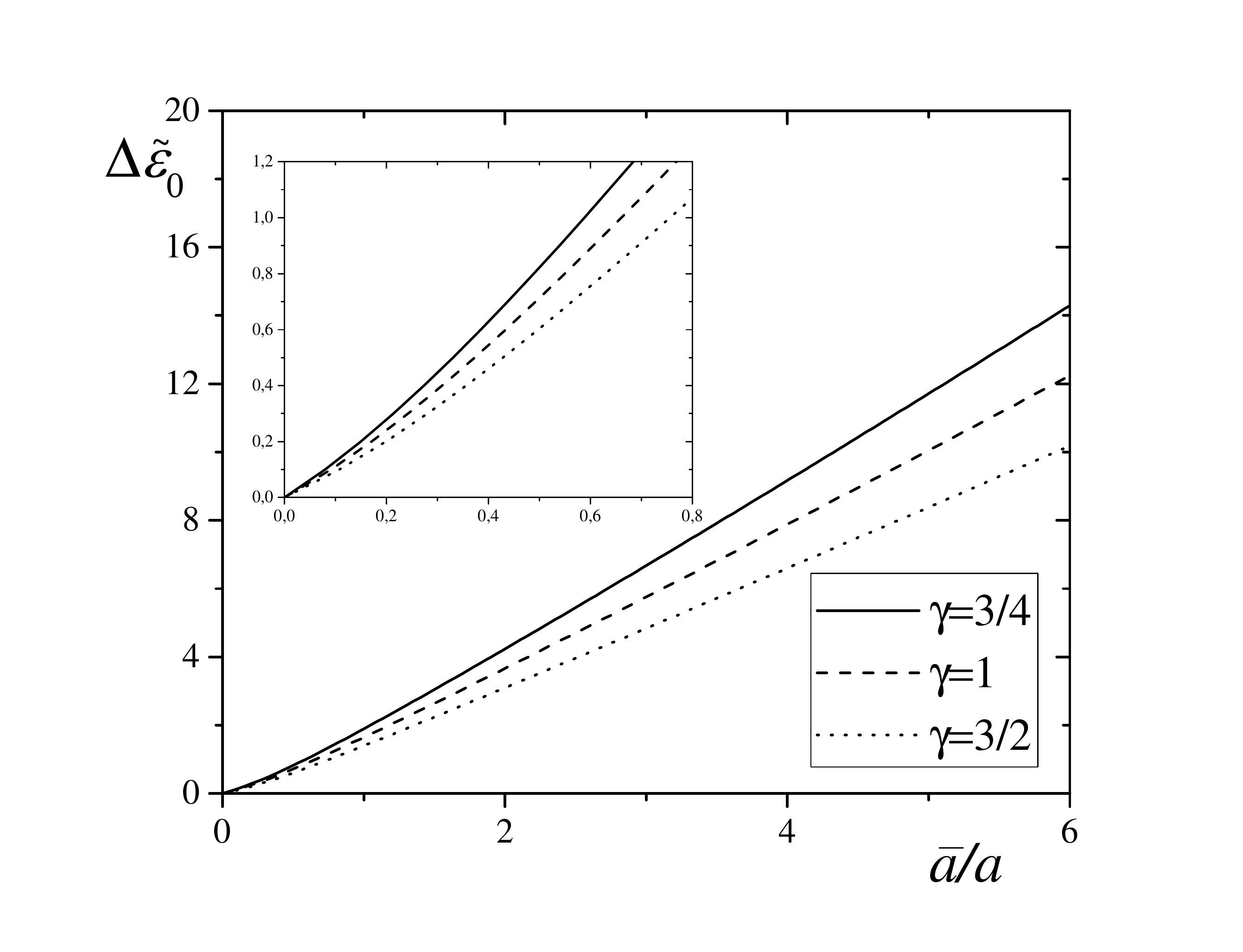}
	\includegraphics[scale=0.27]{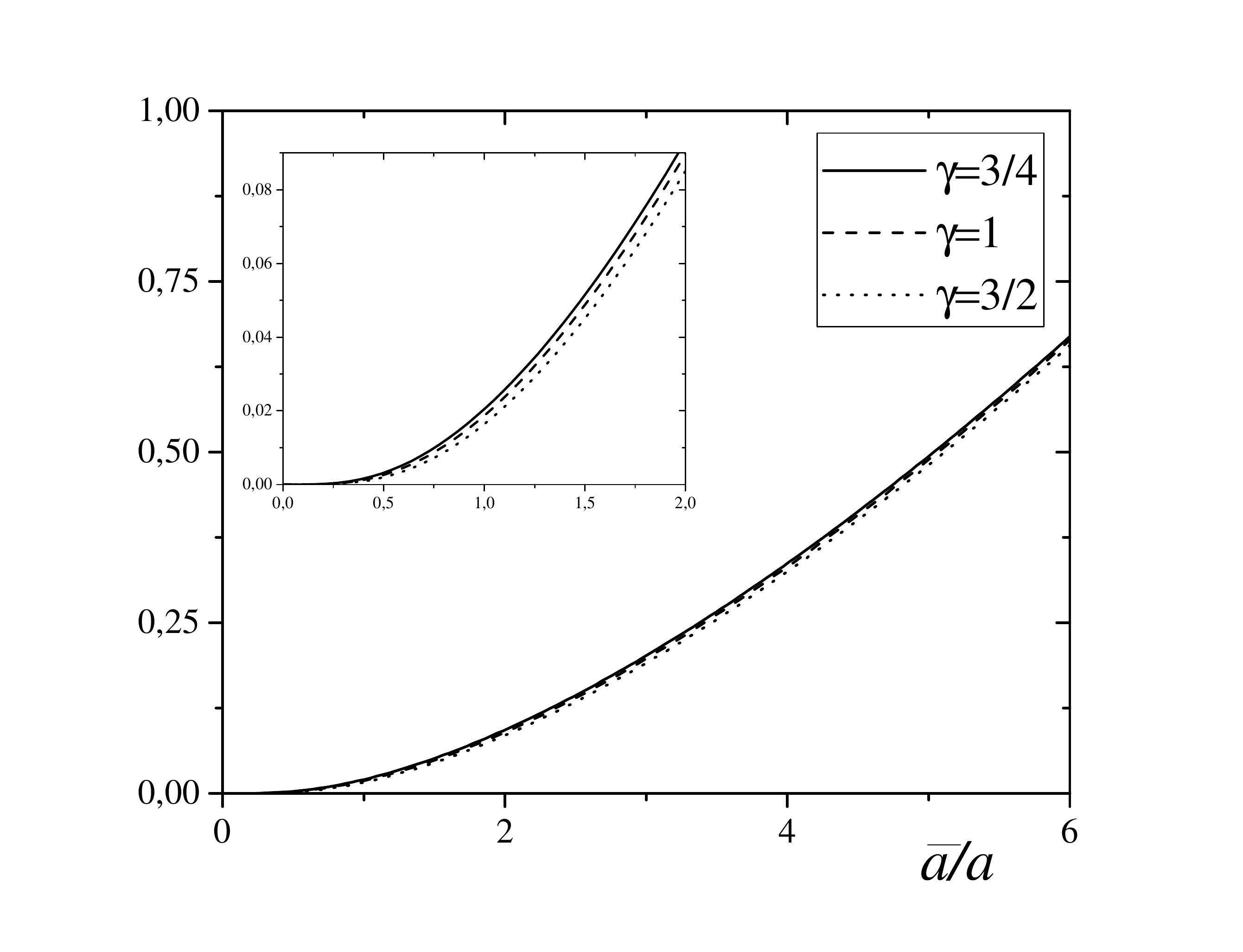}
\vspace{-2mm}
	\caption{The repulsive polaron: left-hand panel is the self-energy of the impurity atom, right-hand panel is the ratio of the spectrum damping to the self-energy impurity atom. (Top row $\rho a^3=5\times10^{-4}$, bottom row $\rho a^3=3\times10^{-2}$). \label{fig1}}
\end{figure}

It is quite natural that the Bose polaron remains stable in the weak-coupling limit but surely the most interesting is the region of strong boson-impurity interaction where the use of the Brillouin-Wigner perturbation theory can give some non-trivial results.
In this limit, the spectrum damping has a square root behavior on the polaron energy
\be
\label{64}
\tilde{\Gamma}_0=\frac{\pi}{4\sqrt{2}}\sqrt{\frac{\rho a^3}{\pi}}\left(\frac{\bar a}{a}\right)^2\sqrt{\frac{1+\gamma}{\gamma}}\sqrt{\Delta\tilde{\varepsilon}_0}\,, \qquad  \Delta\tilde{\varepsilon}_0\gg 1.
\ee
The results of numerical computations of the polaron energy and damping rate $\Gamma_0/\Delta\varepsilon_0$ are depicted in figures~\ref{fig0}, \ref{fig1}.
As it is seen, the attractive polaron can have both positive and negative values of the self-energy (see left-hand panel in figure~\ref{fig0}). The energy of the repulsive polaron  is always positive [see left-hand panel in figure~(\ref{fig1})]. In the case of the attractive interaction between the impurity atom and Bose particles (when $\bar{a}<0$), the heavier impurities rather leave the potential well formed by Bose   environment (see left-hand panel in figure~\ref{fig0}, dot line).
If we described the interaction between Bose particles by a more realistic potential, then the heavy impurities conversely would be deeper in the potential well. In the case of weak and strong repulsive interaction between the impurity atom and Bose particles (when $\bar{a}>0$), the self-energy increases linearly with increasing the parameter $\bar a/a$ (see left-hand panel in figure~\ref{fig1}).
The spectrum of the impurity atom is damped in the case of strong attraction (see  figure~\ref{fig0}, right-hand panel), when the energy of impurity is positive (see  figure~\ref{fig0}, right-hand). In the case of repulsive interaction between the impurity atom and the Bose system, the spectrum is always damped (see  figure~\ref{fig1}, right-hand). Thus, the spectrum damping grows with an increase of the interaction strength and in the case $\rho a^3=5\times10^{-4}$, our calculations indicate that the attractive polaron is well-defined for $\bar a/a<15\div 16$, whereas the damping rate of a repulsive polaron becomes the same order of magnitude as the self-energy for $\bar a/a\simeq 20$ (see right-hand upper panel in figure~\ref{fig0} and right-hand upper panel in figure~\ref{fig1}). When $\rho a^3=3\times10^{-2}$, the condition of the applicability of the presented approach is restricted to $\bar a/a<5\div6$ for the attractive and to $\bar a/a<8\div 10$ for the repulsive polarons, respectively.

\begin{figure}[!b]
\vspace{-4mm}
\centering
	\includegraphics[scale=0.27]{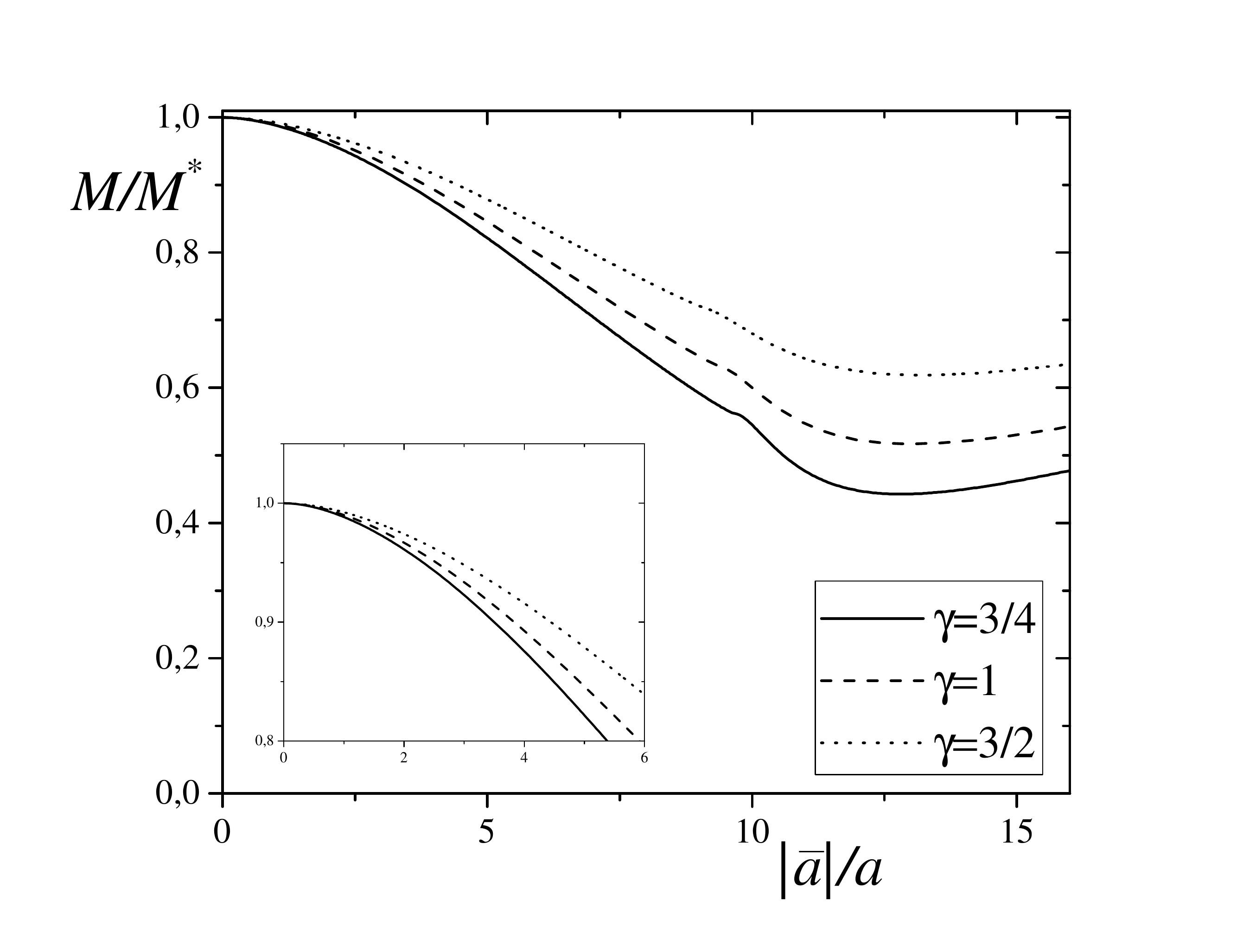}
	\includegraphics[scale=0.27]{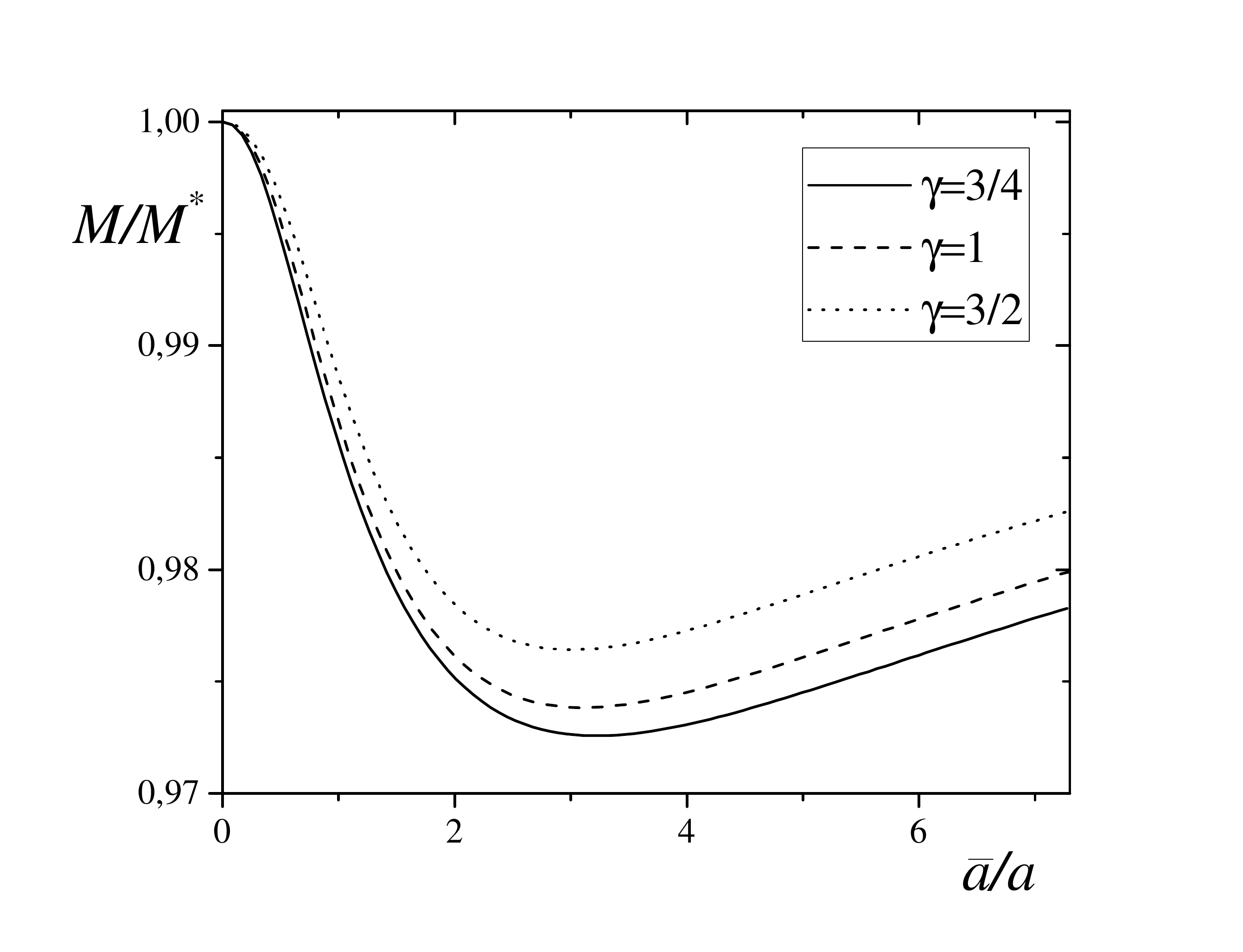}
\vspace{-5mm}\\
	\includegraphics[scale=0.27]{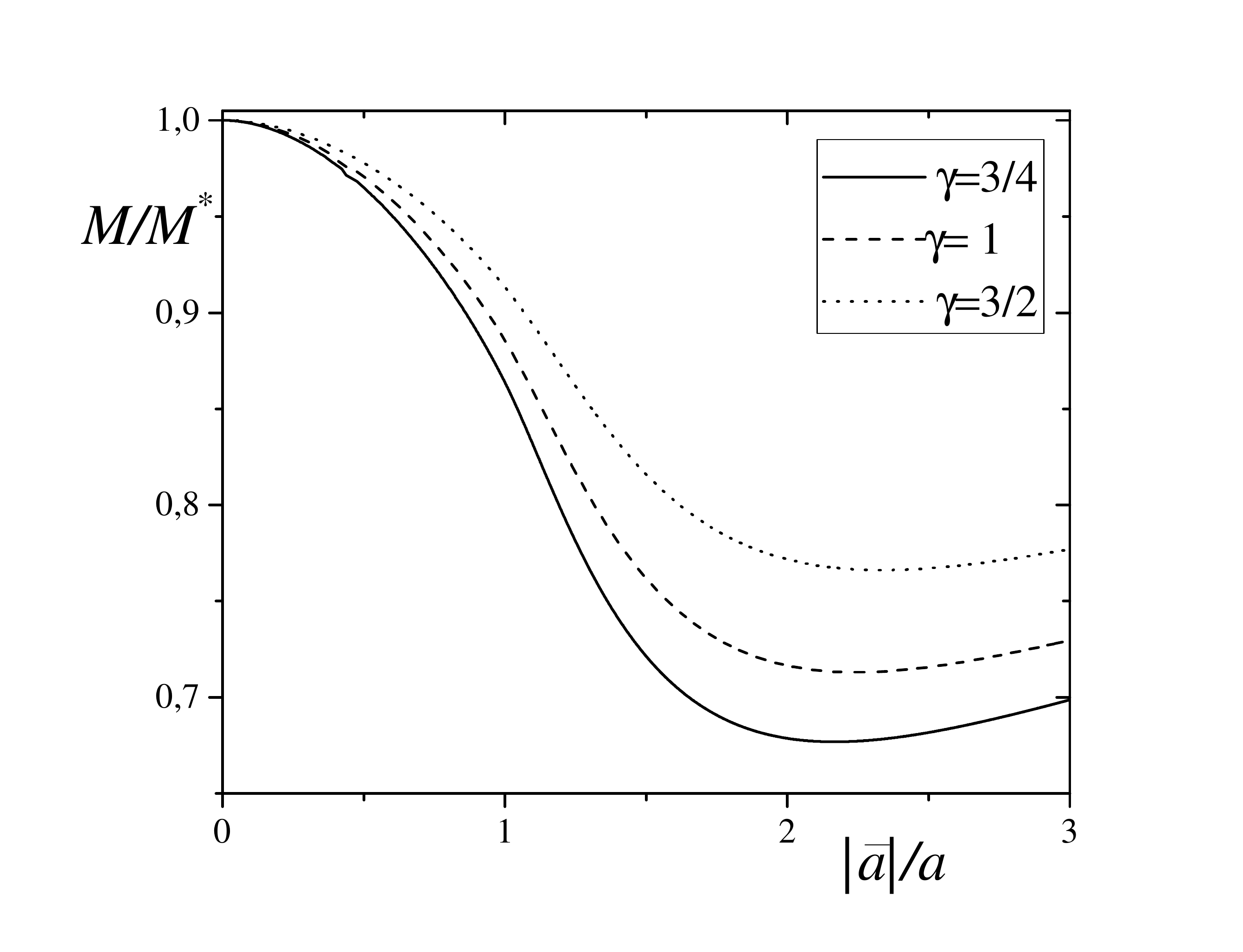}
	\includegraphics[scale=0.27]{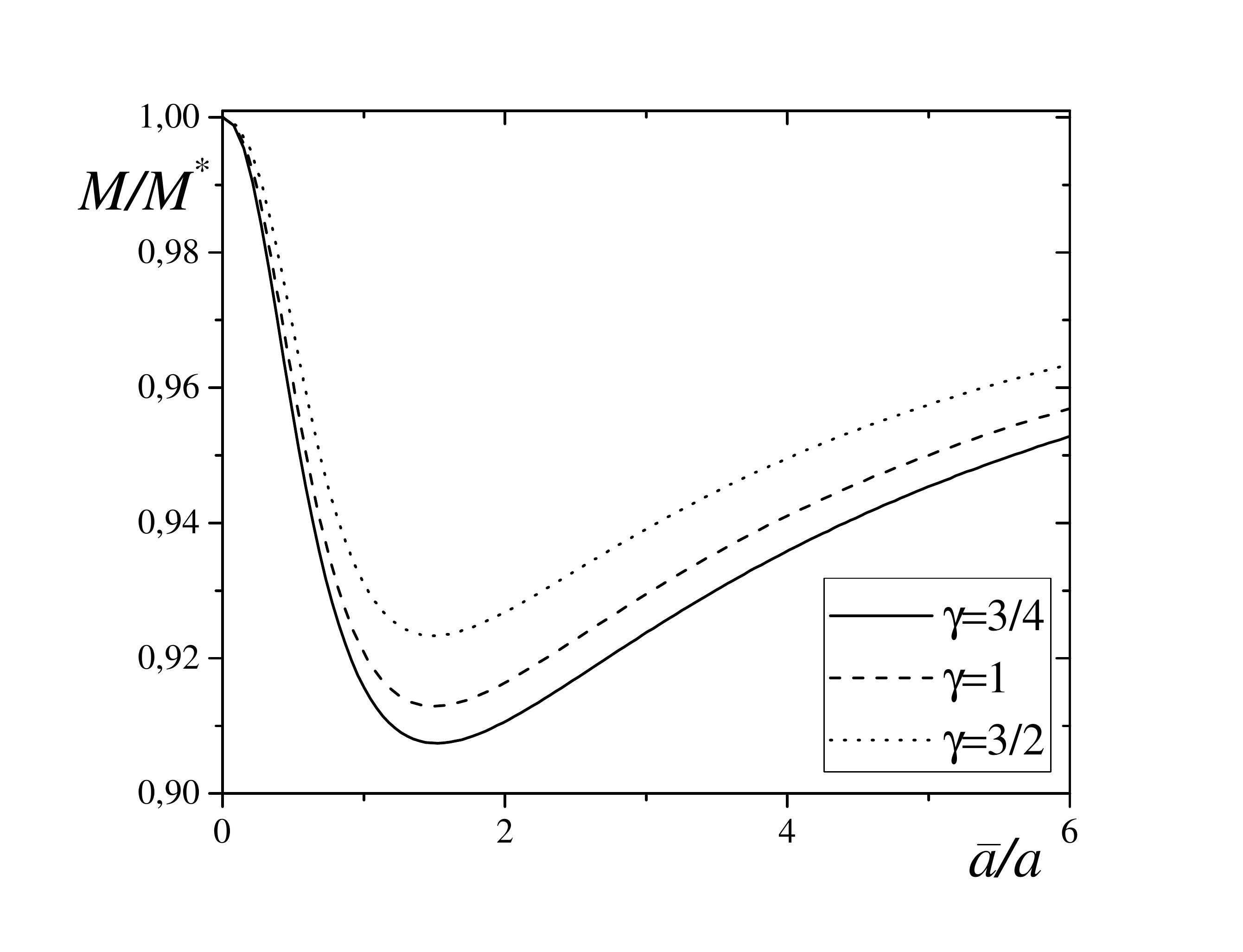}
\vspace{-4mm}
	\caption{The effective mass of the impurity atom: left-hand panel is the attractive polaron, right-hand panel is the repulsive polaron. Top row $\rho a^3=5\times10^{-4}$, bottom row $\rho a^3=3\times10^{-2}$.\label{fig3}}
\end{figure}

Therefore, the calculations of the effective mass were performed in the region where $\Delta\varepsilon_q>\Gamma_q$. The computation of these integrals can be reduced to the previously found one. In particular, contribution (\ref{47}) to the effective mass can be obtained by differentiation of the expression for the self-energy (\ref{45}) with respect to the parameter $\Delta{\varepsilon}_0$. In the thermodynamic limit, it is not difficult to calculate  the sum with a cubic term in (\ref{46}) by differentiation with respect to the parameter $\gamma^{-1}$. Thus, the effective mass in dimensionless units can be written as follows:
\be\label{48}
\frac{M}{M^{*}}=1-\frac{4}{3}\sqrt{\frac{\rho a^3}{\pi}}{\left(\frac{\bar {a}}{a}\right)}^2\frac{(1+\gamma)^2}{\gamma^3}\frac{\partial^2 {\cal{I}}_2(\gamma,\Delta\tilde{\varepsilon}_0)}{\partial (\gamma{^{-1}})^{2}}\left[1+4\sqrt{\frac{\rho a^3}{\pi}}{\left(\frac{\bar {a}}{a}\right)}^2\frac{(1+\gamma)^2}{\gamma}\frac{\partial {\cal{I}}_1(\gamma,\Delta\tilde{\varepsilon}_0)}{\partial \Delta\tilde{\varepsilon}_0}\right]^{-1},
\ee
where ${\cal{I}}_{1,2}(\gamma,\Delta\tilde{\varepsilon}_0)$ is presented in appendix \ref{sec:Apa}.
An increase of the attractive interaction (see figure~\ref{fig3} left-hand panel) and repulsive interaction (see figure~\ref{fig3} right-hand panel) between the impurity and Bose particles leads to an increase of the effective mass of the polaron to a certain critical value of the interaction. In the case of very strong interaction, the effective mass gradually decreases. Note that in the paper \cite{Strez}, the effective mass of the impurity atom $^3$He in the liquid $^4$He with the real interaction potential between Bose particles was found for the first time by means of the Brillouin-Wigner theory. In the paper \cite{Vakarchuk1}, the energy of the moving impurity in the liquid $^4$He was found. The author showed that the interaction potential between the Bose particles and the impurity atom can be expressed as a function of the structure factor of the liquid $^4$He. In the work \cite{Vakarchuk2}, using a deformed Heisenberg algebra, the effective mass and the separation energy of the impurity atom $^3$He for different values of the density of $^4$He is calculated.

\section{Conclusion}
In conclusion, we have analyzed the spectral properties of attractive and repulsive Bose polarons in a dilute Bose condensate. Particulary, using the Rayleigh-Schr\"odinger perturbation theory, we obtained a  full momentum dependence of the energy and damping of the impurity atom which is moving in the Bose gas with a weak short-range repulsion. It is also shown that the Bose polaron behavior obtained in the second-order Brillouin-Wigner perturbation theory qualitatively reproduces the results of a more complicated T-matrix approach \cite{Rath} and variational calculations \cite{Li}.

\appendix
\section{Appendix}\label{sec:Apa}
The corrections to the spectrum of the impurity $\epsilon_p(\gamma)$ (\ref{43}) in the limit of the standard perturbation  theory depend on the mass and momentum of the impurity.

a) The case $\gamma<1$ (the light impurity):
\begin{align}\nonumber
\epsilon_p(\gamma)&=\frac{8}{1-\gamma}\left\{1 -\frac{1}{1-\gamma^2}\frac{1}{p}\int_0^p{\rd}p\,\theta\left({p-\sqrt{1-\gamma^2}}\right)
\Bigg[\frac{\gamma^2-\gamma^4+p^2\left(1+\gamma^2\right)}{2\sqrt{1+p^2-\gamma^2}}\ln{\Bigg|\frac{1+\sqrt{1+p^2-\gamma^2}}{1-\sqrt{1+p^2-\gamma^2}}\Bigg|} \right.\\&\left.
\phantom{\int_0^p}+\gamma p\ln\Big|\frac{\gamma-p}{\gamma+p}\Big|\Bigg]\right\},\label{70}
\end{align}
and the condition, $p>\sqrt{1-\gamma^2}$.

In the limit $p\to 0$ equation~(\ref{70}):
\be\label{71}\nonumber
\epsilon_0(\gamma)=\frac{8}{1-\gamma}\left(1-\frac{\gamma^2}{2\sqrt{1-\gamma^2}}\ln\Bigg|\frac{1+\sqrt{1-\gamma^2}}{1-\sqrt{1-\gamma^2}}\Bigg|\right).
\ee

b) For the case $\gamma>1$ (the heavy impurities):
\begin{align}\label{76}\nonumber
\epsilon_p(\gamma)&=\frac{8}{1-\gamma}\left\{1+\frac{1}{\gamma^2-1}\frac{1}{p}\int_0^p
{\rd}p\,\theta\left(\sqrt{\gamma^2-1}-p\right)
\Bigg[\frac{\gamma^2-\gamma^4+p^2\left(1+\gamma^2\right)}{\sqrt{\gamma^2-1-p^2}}\arctan{\sqrt{\gamma^2-1-p^2}}\right.\\ \nonumber
&\quad+\gamma p \ln\Big|\frac{\gamma-p}{\gamma+p}\Big|\Bigg]
+\frac{1}{\gamma^2-1}\frac{1}{p}\int_0^p{\rd}p\,\theta\left( p-\sqrt{\gamma^2-1}\right)
\Bigg[\frac{\gamma^2-\gamma^4+p^2\left(1+\gamma^2\right)}{2\sqrt{1+p^2-\gamma^2}}\ln\Bigg|\frac{1+\sqrt{1+p^2-\gamma^2}}{1-\sqrt{1+p^2-\gamma^2}}\Bigg|\\
&\left.\phantom{\int_0^p}+\gamma p \ln\bigg|\frac{\gamma-p}{\gamma+p}\bigg|\Bigg]+\frac{\gamma^2}{\sqrt{\gamma^2-1}}\arctan{\sqrt{\gamma^2-1}}
\right\},
\end{align}
the correction (\ref{76}) in the limit of the motionless impurity we obtain:
\be\label{77}\nonumber
\epsilon_0=\frac{8}{1-\gamma}\left(1-\frac{\gamma^2}{\sqrt{\gamma^2-1}}\arctan{\sqrt{\gamma^2-1}}\right).
\ee
c) For the case $\gamma=1$
\be\label{78}
\epsilon_p=\frac{4}{p}\int_0^p\Bigg[\frac{1+p^2}{p^2}-\frac{\left(1-p^2\right)^2}{2p^3}\ln{\Big|\frac{1+p}{1-p}\Big|}\Bigg]{\rd}p,
\ee
when $q\to 0$ in (\ref{78}), we obtain the results of the corrections: $\epsilon_0=\frac{32}{3}$.

The explicit form of the correction $\epsilon_0(\gamma,\Delta\tilde{\varepsilon}_0)$ to the self-energy of the Bose polaron (\ref{481}):
\be\label{79}
\epsilon_0(\gamma,\Delta\tilde{\varepsilon}_0)=\frac{8}{1-\gamma}\left[1+\left(\frac{\gamma^2}{1-\gamma}+\frac{\gamma\Delta\tilde{\varepsilon}_0}{2}\frac{1+\gamma}{1-\gamma}
\right){\cal{I}}_1(\gamma,\Delta\tilde{\varepsilon}_0)-\left(\frac{\gamma^2}{1-\gamma}+\frac{\gamma\Delta\tilde{\varepsilon}_0}{2}\right){\cal{I}}_2(\gamma,\Delta\tilde{\varepsilon}_0)\right];
\ee
and here, the quantities  ${\cal{I}}_{1,2}(\gamma,\Delta\tilde{\varepsilon}_0)$ are the integrals:
\bee\label{80}\nonumber
{\cal{I}}_1(\gamma,\Delta\tilde{\varepsilon}_0)=\int_{-1}^{1}\frac{{\rd}x}{(x^2-x_{-}^2)(x^2-x_{+}^2)}\,,\qquad {\cal{I}}_2(\gamma,\Delta\tilde{\varepsilon}_0)=\int_{-1}^{1}\frac{x^2{\rd}x}{(x^2-x_{-}^2)(x^2-x_{+}^2)};
\eee
here,
\be\label{81}\nonumber
x_{\pm}=\frac{1}{1-\gamma}\Bigg(1+\gamma\Delta\tilde{\varepsilon}_0\pm\sqrt{\left(1+\gamma\Delta\tilde{\varepsilon}_0\right)^2+\gamma^2-1}\Bigg).
\ee
The result of the integration depends on the ratio between the parameters $\gamma$ and $\Delta\tilde{\varepsilon}_0$.

\section*{Acknowledgements}
This work was partly supported by Project FF-30F (No. 0116U001539) from the
Ministry of Education and Science of Ukraine. The authors appreciate the help of Dr. Mykola Stetsko
in the preparation of this article.

One of the authors (I.~Vakarchuk) expresses satisfaction with the fact that he was invited to participate in the publication of articles in honor of Yu.~Holovatch. He has been one of my first students (since his schooldays), who defended his thesis on phase transitions, and is engaged now in the research of complex systems of non-physical nature using methods of statistical physics. I wish him new successes in this extremely interesting area of human activity.

\newpage
\ukrainianpart

\title{Поведінка домішкового атома в слабковзаємодіючому бозе-газі}
\author{Г.~Паночко\refaddr{label1}, В.~Пастухов\refaddr{label2}, I.~Вакарчук\refaddr{label2}}

\addresses{
\addr{label1} Природничий коледж, Львівський національний університет імені Івана Франка, \\ вул. Тарнавського, 107, 79010 Львів, Україна
\addr{label2} Кафедра теоретичної фізики, Львівський національний університет імені Івана Франка,\\ вул. Драгоманова, 12, 79005 Львів, Україна
}

\makeukrtitle

\begin{abstract}
\tolerance=3000%
В роботі вивчено поведінку домішкового атома у розрідженому бозе-конденсаті в границі низьких температур. Отримано пертурбативним методом спектр домішки та його загасання в залежності від імпульсу. В рамках теорії збурень Брілюена–Вігнера знайдено власну енергію притягального та відштовхувального полярона в довгохвильовій границі. Також вивчено питання стабільності полярона в слабковзаємодіючому бозе-газі.
\keywords бозе-полярон, загасання спектра, власна енергія

\end{abstract}
\end{document}